\documentclass[fleqn,10pt]{wlscirep}
\usepackage{bm}
\usepackage{bookmark}
\title{Quantum Transports in Two-Dimensions with Long Range Hopping}

\author[1,2,3]{Si-Si Wang}
\author[4]{Kangkang Li}
\author[1]{Yi-Ming Dai}
\author[1,3]{Hui-Hui Wang}
\author[1]{Yi-Cai Zhang}
\author[1,3,2 *]{Yan-Yang Zhang}

\affil[1]{School of Physics and Materials Science, Guangzhou University, 510006 Guangzhou, China}
\affil[2]{School of Mathematics and Information Science, Guangzhou University, 510006 Guangzhou, China}
\affil[3]{Huangpu Research and Graduate School of Guangzhou University, 510700 Guangzhou, China}
\affil[4]{Department of Physics, Zhejiang Normal University, Jinhua 321004, China}
\affil[*]{yanyang@gzhu.edu.cn}

\begin{abstract}
We investigate the effects of disorder and shielding on quantum transports in a two dimensional system with all-to-all long range hopping. In the weak disorder, cooperative shielding manifests itself as perfect conducting channels identical to those of the short range model, as if the long range hopping does not exist. With increasing disorder, the average and fluctuation of conductance are larger than those in the short range model, since the shielding is effectively broken and therefore long range hopping starts to take effect. Over several orders of disorder strength (until $\sim 10^4$ times of nearest hopping), although the wavefunctions are not fully extended, they are also robustly prevented from being completely localized into a single site. Each wavefunction has several localization centers around the whole sample, thus leading to a fractal dimension remarkably smaller than 2 and also remarkably larger than 0, exhibiting a hybrid feature of localization and delocalization. The size scaling shows that for sufficiently large size and disorder strength, the conductance tends to saturate to a fixed value with the scaling function $\beta\sim 0$, which is also a marginal phase between the typical metal ($\beta>0$) and insulating phase ($\beta<0$).  The all-to-all coupling expels one isolated but extended state far out of the band, whose transport is extremely robust against disorder due to absence of backscattering. The bond current picture of this isolated state shows a quantum version of short circuit through long hopping.
\end{abstract}

\begin{document}

\flushbottom
\maketitle
%
%
\thispagestyle{empty}

\section*{Introduction}

Long range orders and behaviors from short range coupling are among the most important themes of condensed matter physics. Even theoretical models based on nearest neighbor hopping and interaction can be used to describe vast amounts of physical phenomena, including the energy band, magnetism, metal-insulator transition (MIT) and topological states, etc.\cite{Mahan,AdersonTransitionReview1,AdersonTransitionReview2,Hasan2010,XLQiRMP}.
On the other hand, systems with long range coupling have attracted many interests recently. For example, in the photonic lattice, controllable long range coupling can be realized by shaping the spectrum of the optical pump\cite{Bell2017}, or by the optical gain\cite{OpticalGain}. Simulating long range hopping is proposed by periodically driven superconducting qubits\cite{SCQubits}. Rydberg atomic arrays are considered as a promising platform for quantum information application\cite{Rydberg1,Rydberg6}, where the ultra-long coupling between atoms plays a crucial role\cite{Rydberg2,Rydberg3,Rydberg4,Rydberg6,Rydberg9,Rydberg7,Rydberg8,Rydberg5}. Among them, two-dimensional (2D) arrays with long range interactions have been recently fabricated and studied\cite{Rydberg1,Rydberg6,Rydberg5}.

The long range coupling between atoms or molecules can also be realized by the mediation of phonons\cite{Phonon1,Phonon2}, photons\cite{Photon1,Photon2} or an optical cavity\cite{CavityExperiment1,CavityExperiment2,CavityExperiment3} in different physical systems, including conventional electronic systems\cite{CavityExperiment1,CavityElectron2,CavityElectron3}. Another interesting realization of quantum lattice model is the electric circuit network\cite{Circuit1,Circuit2,Circuit4,Circuit3}. Since wires can be connected in arbitrary ways, in principle this system can simulate models with arbitrary ranges of hopping in any spatial dimensions.

Long range coupling systems possess many special and useful properties. In a Rydberg array for quantum computing, the long range coupling can compensate for low fidelity and therefore enables better algorithmic performances\cite{Rydberg6}. A one-dimensional dimerized superconducting circuit lattice with long-range hopping is proposed to be a phase-robust topological router\cite{TopologicalRouter}. Rich phenomena of localization from some long range coupling models have been recently noticed\cite{Localization2019}. Some of them may be rather counterintuitive. For example, although long range hopping seems to greatly enhance connectivity between sites, it does not necessarily result in an enhanced quantum transport\cite{Localization2019,Shielding2016}.

Even for one-dimensional (1D) systems, theoretical studies have shown many novel properties, for example, breaking of ergodicity\cite{Ergodicity}, cooperative shielding in many-body systems\cite{ShieldingManyBody}, weakened topological bulk-boundary correspondence\cite{BulkBoundary}, and subdiffusive phases in an open clean system\cite{Subdiffusive2021}. The measurement-induced phase transitions of long range coupling systems are theoretically investigated\cite{LR2022N1,LR2022N2,LR2022N3,LR2022N4}, with practical applications in qubits. Based on a model with all-to-all and distance-independent hopping\cite{StarModel}, some interesting phenomena in 1D have been predicted recently\cite{Shielding2016,DisorderEnhanced2021,DisorderEnhancedPhysics}. The first one is a cooperative shielding in the single particle picture, i.e., absence of effects from the long range hopping for most of the states in the clean limit\cite{Shielding2016}. The second is the disorder-enhanced and disorder-independent transport, if a large-bias current is considered\cite{DisorderEnhanced2021}.

In this manuscript, we generalize the above all-to-all model to a 2D version, and theoretically investigate its quantum transports. By presenting transport evidences, we find that the cooperative shielding persists in the weak disorder limit. Although a fixed boundary condition breaks the perfect shielding, this breaking will be negligible for a sufficiently large sample. With increasing disorder, the shielding is destroyed and the transports will be remarkably different from the short range counterpart. We reveal the microscopic pictures of these transports by showing the real space distributions of bond currents, so that the roles of bonds with different ranges can be seen vividly. Furthermore the localization property is discussed by calculating the fractal dimension of eigenstates, and by performing size scaling of the conductance. Over several orders of strong disorder, most states exhibit a hybrid feature from (or a marginal feature between) localization and delocalization. A unique feature of this model is the existence of a single isolated state far away from the band states\cite{StarModel,Shielding2016}, with a large gap proportional to the size of the sample. We find it is an extended isolated state with very robust transport. The physical origins are also discussed.

\section*{ Model and Method}

\begin{figure}[htbp]
\centering
\includegraphics*[width=0.5\textwidth,bb=10 173 670 735]{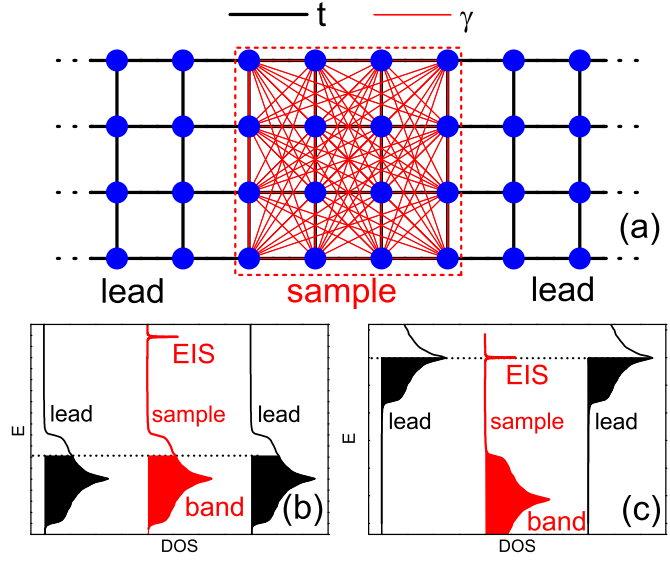}
\caption{(Color online) Illustration of transport setups in our simulation. (a) Two-terminal conductance setup, where a $4 \times 4$ sample (encircled by the red dashed-line square) with long range hopping is connected to two leads with short range hopping. Black (red) bonds represent the nearest (long range) hopping $t$ ($\gamma$). (b) and (c) illustrate the energy band configurations among the sample (red) and leads (black) used in Sections III and IV, respectively, where the black dashed line labels the chemical potential.
Created from OriginPro 8 SR0 (URL: http://www.OriginLab.com).}
\label{FigIllustration}
\end{figure}

Our 2D model is a generalization from the 1D counterpart\cite{Shielding2016,DisorderEnhanced2021},
which is illustrated as the sample enclosed by the red dashed-line square in Figure \ref{FigIllustration} (a).
It is defined on a square lattice with the spinless Hamiltonian
\begin{eqnarray}
H_{\mathrm{LR}}&=&H_{\mathrm{NN}}+H_{\mathrm{AA}}\nonumber \\
&\equiv &\sum_{\langle ij \rangle}t c^{\dagger}_{i}c_{j}+\sum_{ ij}\gamma c^{\dagger}_{i}c_{j} \label{EqHamiltonian}
\end{eqnarray}
where $c^{\dagger}_{i}$ ($c_{i}$) creates (annihilate) an electron at site $i$.
Here $H_{\mathrm{NN}}$ contains the conventional nearest neighbor hopping with the magnitude $t$ [black bonds in Figure \ref{FigIllustration} (a)].
The second term, $H_{\mathrm{AA}}$ includes all-to-all and distance independent long range hopping $\gamma$ [red bonds in Figure \ref{FigIllustration} (a)], which can be realized by a cavity-assisted technology\cite{DisorderEnhanced2021,CavityExperiment1,CavityExperiment2}.
It has been argued that this distance independent long range hopping $\gamma$
grasps the main physics arising from the coupling of the molecules
with the cavity mode, since the coupling to the cavity mode is the same
for all molecules\cite{DisorderEnhanced2021}.
Throughout this paper, $t=1$ will be used as the energy unit, and $\gamma=1/2$ is identical to previous 1D counterparts\cite{Shielding2016,DisorderEnhanced2021}.
For simplicity, we adopt the sign convention that $t,\gamma >0$.
In $H_{\mathrm{AA}}$, the diagonal terms $\gamma c^{\dagger}_{i}c_{i}$ are intentionally included.
This global and trivial energy shift makes the band center of $H_{\mathrm{LR}}$ (referred as the ``long range model'' hereafter) identical to that of $H_{\mathrm{NN}}$ (referred as the ``short range model'' hereafter), which will be convenient in following calculations.

For a finite sample with $N\equiv N_x \times N_y$ lattice sites, and with periodic boundary condition for $H_{\mathrm{NN}}$ in both directions (there is no difference of boundary conditions for $H_{\mathrm{AA}}$), it can be easily verified that two terms of $H_{\mathrm{LR}}$ are commutable and therefore,
\begin{equation}
C_H\equiv \big[ H_{\mathrm{NN}},H_{\mathrm{LR}} \big] =0.\label{EqCommute}
\end{equation}
As a result, $H_{\mathrm{LR}}$ and $H_{\mathrm{NN}}$ have common eigenfunctions\cite{Shielding2016}. Furthermore, the operator $H_{\mathrm{AA}}$ has $N-1$ degenerate eigenvalues $0$ and one eigenvalue $N\gamma$.  Therefore similar to the 1D case\cite{Shielding2016}, the eigenvalues of the long range model $H_{\mathrm{LR}}$ are also identical to those of the short range model $H_{\mathrm{NN}}$, i.e.,
\begin{equation}
E^{\mathrm{LR}}_n=E^{\mathrm{NN}}_n, \quad 1\leq n \leq N-1,\label{EqIdentical}
\end{equation}
except the highest one (the lowest one if $\gamma<0$), which is
\begin{equation}
E^{\mathrm{LR}}_N=4t+ N \gamma \label{EqIsolatedState}
\end{equation}
for $H_{\mathrm{LR}}$. Here and throughout this manuscript, we always index eigenstates of a Hamiltonian in the ascending order of eigenvalues. Notice from Equation (\ref{EqIsolatedState}) that $E^{\mathrm{LR}}_N$ is size dependent, and this single state is isolated from the energy band consisting of the rest $N-1$ eigenstates. Thus it will be called the isolated state. Correspondingly, the rest $N-1$ eigenstates [distributed within $ [-4t,4t)$] will be called the band, which are simply identical to those of the single band of the short range model [Equation (\ref{EqIdentical})]. We can define the subspace $S_N$ (the isolated state) as the one spanned by this isolated state $|\psi_N\rangle$, and the subspace $S_{N-1}$ (the band) as the one spanned by the rest $N-1$ eigenstates ${|\psi_i\rangle}$. Due to the large gap $\Delta \sim N_x N_y \gamma$, these two subspaces are barely mixed when the sample size is sufficiently large and/or the disorder strength is not strong\cite{Shielding2016}. These novel mathematical structures lead to interesting consequences. For example, because of the above mentioned commutability and unmixing between $S_N$ and $S_{N-1}$, the dynamics within $S_{N-1}$ is shielded from long-range hopping, namely it behaves as if long-range hopping does not exist.
This is the the cooperative shielding, a counterintuitive phenomenon found in the 1D counterpart\cite{Shielding2016}. In the following, we will see that this cooperative shielding is also manifested in 2D quantum transports.

One of our focus is disorder, which is simply included by adding a random onsite potential as
\begin{equation}
V=\sum_{i}W\cdot U_i c^{\dagger}_{i}c_{i},\label{EqDisorderPotential}
\end{equation}
where $U_i$ are independent random numbers uniformly distributed in $(-1/2,1/2)$ and $W$ is the single parameter to characterize the disorder strength. With nonzero $W$, the short range part $H_{\mathrm{NN}}+V$ will not commute with the all-to-all part $H_{\mathrm{AA}}$ again. The shielding and corresponding transport phenomena will be one of the primary themes of this work.

Now let us briefly describe the main methods of calculation. At zero temperature, the two-terminal conductance $G$ of a finite sample is proportional to the transmission (Landauer formula)\cite{Landauer}, and
can be expressed by Green's functions as\cite{Datta,Datta2}
\begin{equation}
G=\frac{e^{2}}{h}\mathrm{Tr}\left[  \Gamma_{L}G^{r}\Gamma
_{R}G^{a}\right]  , \label{EqConductance}%
\end{equation}
where
$G^{r/a}(E)\equiv \left(E\pm -H -\Sigma^{r/a}_L-\Sigma^{r/a}_R\right)^{-1}$ is the dressed retarded/advanced Green's function of the central sample, and $\Gamma_{L(R)}%
=i(\Sigma_{L(R)}^{r}-\Sigma_{L(R)}^{a})$ with $\Sigma_{L(R)}%
^{r/a}$ being retarded/advanced self energies due to the left (right) lead, respectively (The spin degeneracy factor 2 is omitted in this manuscript). The lead self energy is defined as
\begin{equation}
\Sigma=\tau g(E) \tau^{\dagger}, \label{EqSelfEnergy}
\end{equation}
where $g(E)$ is surface Green's function of the semi-infinite lead, and $\tau$ is the coupling Hamiltonian from the sample to the lead\cite{Datta}. Numerically this self energy can be conveniently calculated from a direct diagonalization method\cite{DHLee}. In our calculations, as illustrated in Figure \ref{FigIllustration} (a), we take both leads to be semi-infinite square lattices with only the nearest hopping Hamiltonian $H_{\mathrm{NN}}$, since otherwise, it is numerically inaccessible to calculate the self energy of a semi-infinite lattice with long range hopping.

The local current from site $i$ to $j$ along the bond is\cite{Datta2,Jiang2009}
\begin{equation}
J_{i\rightarrow j}=\frac{2e^2}{h}\mathrm{Im}\big[H_{ij}G^{n}_{ji}\big](V_L-V_R),\label{EqLocalCurrent}
\end{equation}
where $H_{ij}$ the matrix element of the bare Hamiltonian, and $G^{n}=G^r\Gamma_LG^a$ is the correlation function.
Since it is defined in the linear response regime, we simply take the voltage difference $V_L-V_R$ between the left (source) and right (drain) leads to be unity.
In this case, the net current through any transverse cross section of the
sample is numerically equal to the conductance calculated from Equation (\ref{EqConductance}).

Notice that all Green's functions and self energies appearing in Eqs. (\ref{EqConductance}) and (\ref{EqLocalCurrent}) are energy dependent. In the weak disorder regime, the energies in leads are adopted to vary with that of the sample as illustrated in Figure \ref{FigIllustration} This is the case in Figure \ref{FigW0}, Figure \ref{FigW} and Figure \ref{FigCurrent} (a-e). Such a setup can minimize the contact resistance due to the mode mismatch between the sample-lead interface. In the regimes of strong disorder or the isolated state, the energies in leads are fixed at $E=0$ to provide the maximum number of conductive channels (which is equal to $N_y$). This is applied in Figure \ref{FigCurrent} (f), Figure \ref{FigScalingV2} and Figure \ref{FigIsolatedAll}. In these cases, the interface scattering is not an issue because the impurity scattering in the sample is dominating.

\section*{The Band}

\subsection*{A. Shielding at Weak Disorder}

\begin{figure}[htbp]
\centering
\includegraphics*[width=0.65\textwidth,bb=0 250 480 596]{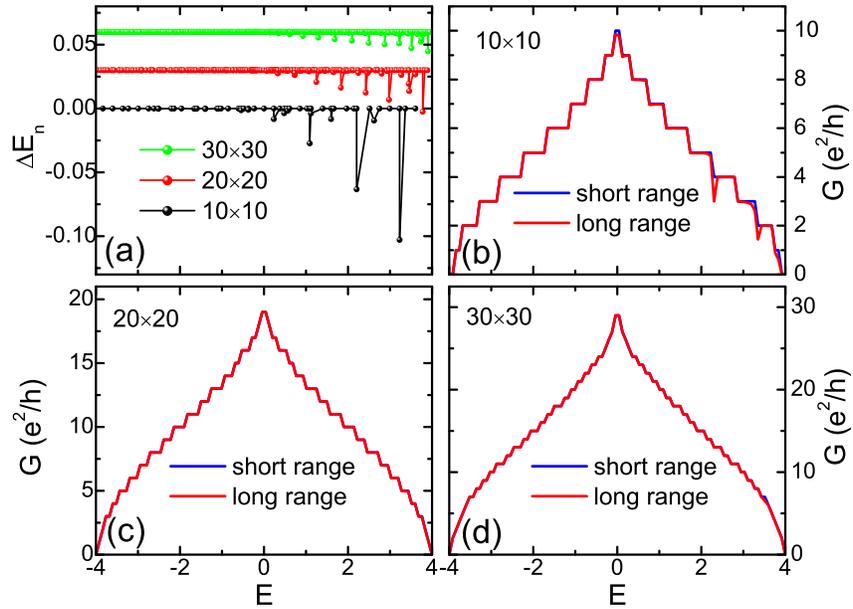}
\caption{(Color online) Results for zero disorder and fixed boundary condition. (a) Difference of eigenergies $\Delta E_n\equiv E^{\mathrm{LR}}_n - E^{\mathrm{NN}}_n$ between the long range and short range models, for a finite sample with sizes $10 \times 10 $ (black), $20 \times 20$ (red) and $30 \times 30$ (green) respectively. The latter two curves are shifted vertically for visual clarity. (b)-(d): the two-terminal conductance $G$ as a function of Fermi energy $E$, for a short range (blue curve) and long range model (red curve), with sample size $10 \times 10$ (b), $20 \times 20$ (c) and $30 \times 30$ (d). Most of the blue curves have been covered by red curves due to an almost perfect match. The transport setup is shown in Figure \ref{FigIllustration} (a) and (b). }
\label{FigW0}
\end{figure}

Let us first investigate properties in the clean limit, $W=0$. In the following calculations, all samples are square shaped with size $N = N_x \times N_x$ to exhibit the 2D nature. The commutator $C_H$ in Equation (\ref{EqCommute}) holds rigorously for periodic boundary conditions, for which a perfect shielding is expected. However, such a torus geometry is not physically applicable for a realistic 2D sheet, especially when it will be connected to conducting leads. Therefore in the rest of our work, we will employ fixed (hard-wall) boundary conditions in both directions for the $H_{\mathrm{NN}}$ term (there is no boundary effects for the $H_{\mathrm{AA}}$ term). In this case, the commutator matrix $C_H$ will not be identically zero. However, we find that nonzero matrix elements only appear when they are associated with boundary bonds of $H_{\mathrm{NN}}$, which only constitute an extremely small portion of the matrix $C_H$. With increasing size, this portion will be even smaller due to the shrinking of boundary-bulk proportion, so the perfect shielding is expected to recover.

We first check the validity of Equation (\ref{EqIdentical}) for isolated samples without being attached to any leads. In Figure \ref{FigW0} (a), we plot the difference of eigenvalues from the long range and short range models, without disorder and in the presence of fixed boundary condition, for three different sample sizes. Notice the latter two curves are shifted vertically for better visual clarity. One can see that remarkable differences mostly occur in the high energy region near the band top (and also near the isolated state), and the difference shrinks with increasing sample size. Therefore it is reasonable to expect a perfect shielding effect recovers for sufficiently large sample.

As an example of physical observable, the two-terminal conductance $G$ of a sample connected with leads (short range model) will be calculated by using Equation (\ref{EqConductance}). Notice that all Green's functions and self energies appearing in Equation (\ref{EqConductance}) are energy dependent. Here, as illustrated in Figure \ref{FigIllustration} (b), we adopt a uniform Fermi energy $E$ among the central sample and leads, i.e., $G(E)=\frac{e^{2}}{h}\mathrm{Tr}\left[  \Gamma_{L}(E)G^{r}(E)\Gamma
_{R}(E)G^{a}(E)\right]$, so that wavefunctions in these regions can have a best mode match, which is expected from the perfect cooperative shielding.

The resulting $G$ of the long range model sample as a function of Fermi energy $E$ is presented as red curves in Figure \ref{FigW0} (b)-(d), for different sample sizes. For comparison, the result for a sample replaced by a short range model ($\gamma=0$) with the same size is also plotted as blue curves in each panel. For the latter case, since the sample and leads are completely identical and perfectly matched, the transmission is therefore perfectly quantized as typical integer steps\cite{Datta}. Then, it is interesting to see that the result from the former case also matches these quantization steps nicely. Distinguishable differences only occur occasionally in the high energy region in Figure \ref{FigW0} (b), i.e., for the smallest sample.

To understand these results, we should remember that our simulation of quantum transports is based on a fully coherent picture, which makes the results highly sensitive to any imperfection or mismatch in the structure\cite{Datta2}. Even without any disorder in the sample, the mismatch between the sample and leads is sufficient to introduce remarkable scattering at interfaces and therefore destroy the conductance quantization remarkably. For example, a recent study on a low-dimensional structure with long range hopping shows that such a mismatch can even cause a subdiffusive transport through the whole clean sample, instead of a ballistic one\cite{Subdiffusive2021}.
That is to say, in the coherent limit, the scattering at the interface alone can be strong enough to alter the transport property qualitatively. Therefore here, the perfect quantization of coherent transmission through a \emph{heterostructure} [as shown in Figure \ref{FigIllustration} (a)] is a nontrivial phenomenon, which implies the absence of scattering in the sample \emph{and} at the lead-sample interfaces, or a perfect match of wavefunction modes between the sample (long rang model) and leads (short range model), despite the apparent lattice mismatch. In other words, the electron travels as if the long range
hopping did not exist, which is a manifestation of the cooperative shielding\cite{Shielding2016} in 2D quantum transport, when disorder is absent. This shielding is a direct result of Eqs. (\ref{EqCommute}) and (\ref{EqIdentical}). Small deviations from perfect shielding in transports [Figure \ref{FigW0} (b)] are a consequence of the boundary effect (which is not periodic), and will practically vanish for larger sample sizes owing to weaker boundary effects.

\begin{figure*} [htbp]
\begin{center}
\includegraphics*[width=0.97\textwidth,bb=0 129 475 307]{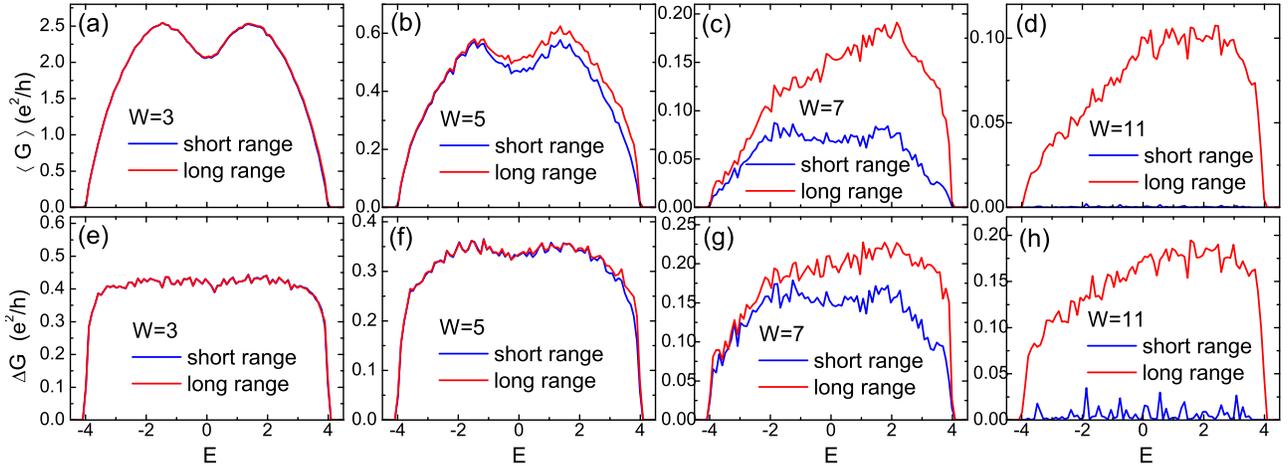}
\end{center}
\caption{(Color online) The disorder averaged conductance $\langle G \rangle $ (first row) and its standard deviation $\Delta G$ (second row) as functions of Fermi energy $E$, for different disorder strengths $W$ as shown. The blue (red) curve is for the short (long) range model. The sample size is $40 \times 40$ and each data point is an average over 1000 disorder samples. All energies are in units of $t$.} \label{FigW}
\end{figure*}

We have seen that in the clean limit, the quantum transport through the 2D long range model is identical to that in the short range model, which is a manifestation of the perfect shielding in 2D. Now we will investigate the effect from disorder, $W\neq0$. Disorder breaks Equation (\ref{EqCommute}) and thus may also break the perfect shielding. In Figure \ref{FigW}, we present the disorder averaged conductance $\langle G \rangle$ (first row) and its standard deviation $\Delta G$ (second row) as functions of the Fermi energy $E$, for different disorder strengths $W$. When $W=3$ [panels (a) and (e)], similar to the clean limit, the results for the short range (blue curves) and the long range (red curves) models are identical. This suggests that the perfect shielding practically survives through weak disorder, which was also observed in 1D \cite{Shielding2016}.

With increasing disorder, for example, $W=5$ [Figure \ref{FigW} (b) and (f)], the results for short and long range models start to deviate gradually. In the case of stronger disorder, $W=7,\, 11$, both the conductance and its fluctuation of the long range model are remarkably larger than those of the short range counterpart.
The existence of long range hopping gives rise to better transport (i.e., larger conductance) in the strong disorder limit.

\begin{figure*} [htbp]
\begin{center}
\includegraphics*[width=0.72\textwidth]{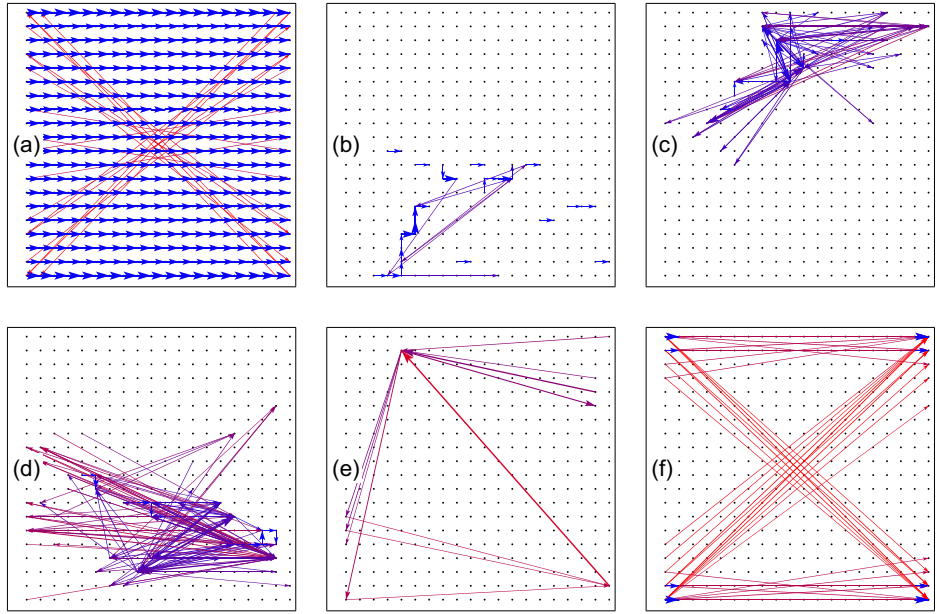}
\end{center}
\caption{(Color online) Distribution of bond currents for a certain disorder configuration on a $20 \times 20$ lattice, with the source (drain) lead connected to the left (right) boundary. (a)-(e): Fermi energy $E=1$ in the band, with disorder strength $W=0,\, 3,\, 5,\, 7,\, 11$, respectively. (f): Sample Fermi energy $E=203.71$ at the EIS, with $W=7$, discussed in Section IV. Each arrow connecting two sites represents the corresponding bond (thus its length indicating the bond length), with the size of the \emph{arrowhead} alone to indicate the magnitude of its current. For visual clarity, only currents with $J \ge 0.4J_{\mathrm{max}}$ are displayed, and long (short) bonds are plotted in red (blue) color, while intermediately long bonds in purple color. } \label{FigCurrent}
\end{figure*}

In order to have a microscopic understanding of these quantum transports, for example, contributions from short and long range hoppings, we present distributions of bond currents associated with some typical regimes in Figure \ref{FigCurrent}. Due to the all-to-all nature of the hopping, there are too many ($\sim (N_x N_y)^2$) bonds with varying lengths intersecting each other, and densely distributed on the lattice. This makes it difficult to present a full and visually distinguishable picture of \emph{all} bond currents, as in conventional short range models\cite{Jiang2009}. Therefore for the purpose of displaying the dominating physics clearly, we adopt some technical tactics in plotting Figure \ref{FigCurrent}. Firstly, a small sample with $20 \times 20$ sites is used. This is smaller than those used in Figure \ref{FigW}, but we have checked (but not shown here) that the physics is identical. Secondly, only currents with magnitudes $J \ge 0.4J_{\mathrm{max}}$ are displayed, where $J_{\mathrm{max}}$ is the maximum magnitude of bond current in this sample. Therefore the displayed currents do not obey the current conservation. Thirdly, the magnitude of a bond current is represented by the size of the \emph{arrowhead} only, while the size of the arrow shaft still represents the real size of the bond connecting two sites. For a better visual clarity, longer (shorter) bonds are plotted in red (blue) color. We do so because, besides the current flowing on it, the actual position and length of each bond are also important information.

Figure \ref{FigCurrent} (a) is the result for zero disorder, corresponding to the case shown in Figure \ref{FigW0} (c). It can be seen that now the currents are almost uniformly carried by nearest bonds (blue arrows), and contributions from most long range bonds are small. This is a clear picture of the perfect shielding: long range coupling hardly plays a role as a result of a delicate quantum coherent effect. Small currents along a few long range bonds (red arrows with small arrowheads) can be attributed to resonant states between hard wall boundaries.

In the presence of very weak disorder, Figure \ref{FigCurrent} (b), the current distribution is also disordered, and contributions from long range bonds start to increase. We notice that this corresponds to the transport shown in
in Figure \ref{FigW} (a) and (e), where the conductances from the short range and long range models still match well. In other words, although the transports are carried by short range and long range bonds, the total current is still very close to that of the short range model. This is another subtle manifestation of the word, the \emph{cooperative} shielding\cite{Shielding2016}. With larger disorder, as shown in Figure \ref{FigCurrent} (c) and (d), the currents distribute in a more chaotic pattern, with long range bonds trying to connect localization centers of the wavefunction.

With these pictures, now we can understand more about Figure \ref{FigW}. When disorder is nonzero but still weak ($W>3$), although these two subspaces are still shielded from each other, long range hops start to manifest themselves by contributing currents and therefore enhancing the conductance. We call it a regime of unperfect shielding, compared to the perfect one (zero effect from long range hopping) for $W=0$ as shown in Figure \ref{FigW0} (c,d) and Figure \ref{FigCurrent} (a). In short range models of localization, it was recently found that microscopically, the dominating transport path can be pinned within a certain range of model parameters (e.g., Fermi energy and disorder potential) \cite{Glassy2019}, since endeavors must be made to find another continuous path composed of sequential short range bonds. However here, thanks to the all-to-all connectivity, an entirely new path can be found more easily upon parameter changes, so that the pinning effect is reduced. Therefore, the transport will be more mutable and sensitive with the change of model parameters, e.g., the disorder potential. This leads to larger fluctuations of the conductance at stronger disorder displayed in Figure \ref{FigW} (g) and (h).

\subsection*{B. Localization Properties at Strong Disorder}

\begin{figure}[htbp]
\centering
\includegraphics*[width=0.9\textwidth,bb=0 139 570 429]{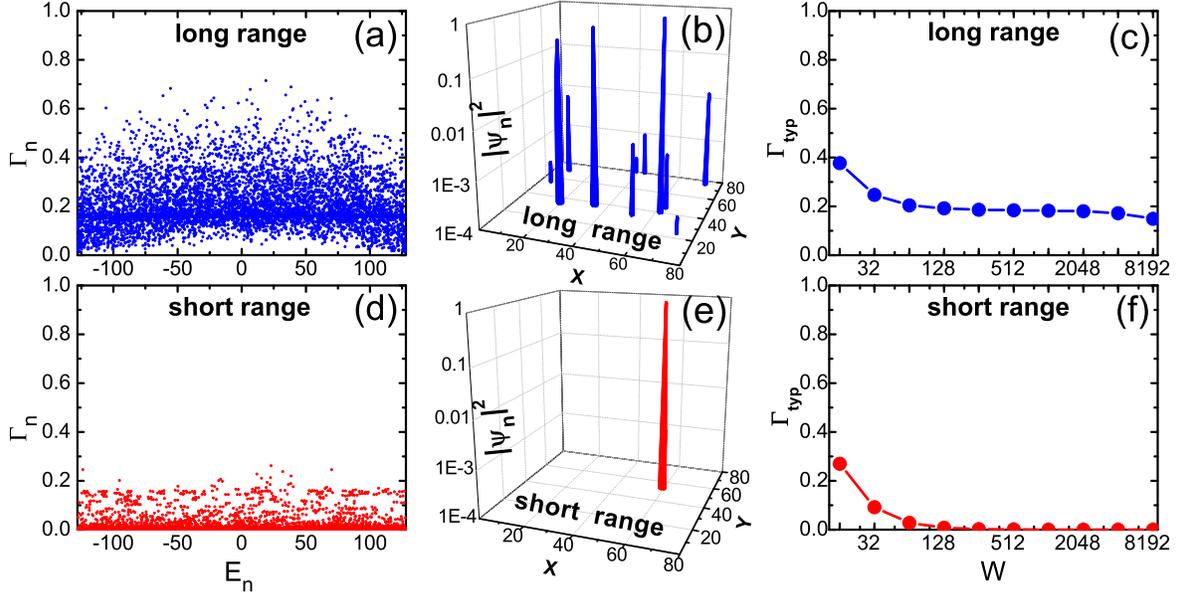}
\caption{(Color online) Eigenstates of a $80 \times 80$ sample for the long range model $\gamma=0.5$ (upper row) and the short range model $\gamma=0$ (lower row). Left column: the fractal dimension $\Gamma_n$ versus eigenenergy $E_n$, with the disorder strength $W=256$. Middle column: the spatial distribution of the squared magnitude of an eigenstate $|\psi_n|^2$ with $\Gamma_n\sim \Gamma_{\mathrm{typ}}$. Right column: $\Gamma_{\mathrm{typ}}$ as a function of the disorder strength $W$ (logarithm scale).  }
\label{FigGammaEig}
\end{figure}

The above discussions were focused on the shielding effect, with rather weak disorder strength $W\lesssim 10$. In the 1D case, the long range model exhibits surprisingly rich behaviors of transports over several orders of the disorder strength $W$\cite{DisorderEnhanced2021}, where all states of short range models should have been completely localized. To have a first observation of our 2D long model with stronger disorder, we now turn to investigate eigenstates of an finite sample detached from any conducting leads. The localization property of the $n$-th eigenstate $\psi_n$ of the sample can be characterized by its fractal dimension defined as
\begin{equation}
\Gamma_n=\frac{\ln (\mathrm{IPR})_n}{\ln N_x},\label{EqFractalDimension}
\end{equation}
where $(\mathrm{IPR})_n$ is the inverse participation ratio $(\mathrm{IPR})_n \equiv \sum_{i}|\psi_{n,i}|^4$ \cite{AdersonTransitionReview2,MobilityEdge,QBZeng2022}. In a 2D system, an extended state corresponds to $\Gamma_n \sim 2$ while a localized state corresponds to $\Gamma_n \sim 0$.

In the left column of Figure \ref{FigGammaEig}, the fractal dimensions $\Gamma_n$ and eigenenergies $E_n$ of band states of a $80\times 80$ sample are plotted, for a certain disorder configuration at $W=256$, with panels (a) and (d) corresponding to the long range and short range models respectively. For the short range model shown in panel (b), as expected, most eigenstates are localized. In fact, the typical value of the fractal dimension
\begin{equation}
\Gamma_{\mathrm{typ}}=\exp(\frac{1}{N_x N_y}\sum_{n=1}^{N_x N_y}\ln\Gamma_{n}),\label{EqGammaTyp}
\end{equation}
for data displayed in Figure \ref{FigGammaEig} (d) is $0.00209\sim 0$. This vanishing dimension suggests that most eigenstates have been localized to a single site, which is verified from the spatial distribution of a typical wavefunction shown in Figure \ref{FigGammaEig} (e). Figure \ref{FigGammaEig} (f) shows $\Gamma_{\mathrm{typ}}$ as a function of $W$, from which we can see that such a single-site localization has been realized when $W\gtrsim 100$.

On the other hand, the behaviors of the long range model are remarkably different.
In Figure \ref{FigGammaEig} (a), most eigenstates possess a significantly nonzero fractal dimension $\Gamma_n$, with a typical value $\Gamma_{\mathrm{typ}}=0.18588$. The spatial distribution of a typical eigenstate in Figure \ref{FigGammaEig} (b) exhibits several  localization centers [instead of one single peak in the short range model shown in panel (e)] which can be connected by long range hoppings. This is similar to that in the case of 1D long range  model where the wavefunction has an extended tail\cite{DisorderEnhanced2021}. In other words, although the disorder breaks the fully extended nature of the wavefunction, the long range hoppings prevent it from being localized into a single site, even at very strong disorder. From the dependence of $\Gamma_{\mathrm{typ}}$ on $W$ in Figure \ref{FigGammaEig} (c), one can see that such a robustness against a single-site localization (and even the value of $\Gamma_{\mathrm{typ}}$ itself)
persists over several orders of disorder strength $W$ (Notice the logarithmic scale of $W$).

\begin{figure}[htbp]
\centering
\includegraphics*[width=0.97\textwidth,bb=0 123 562 359]{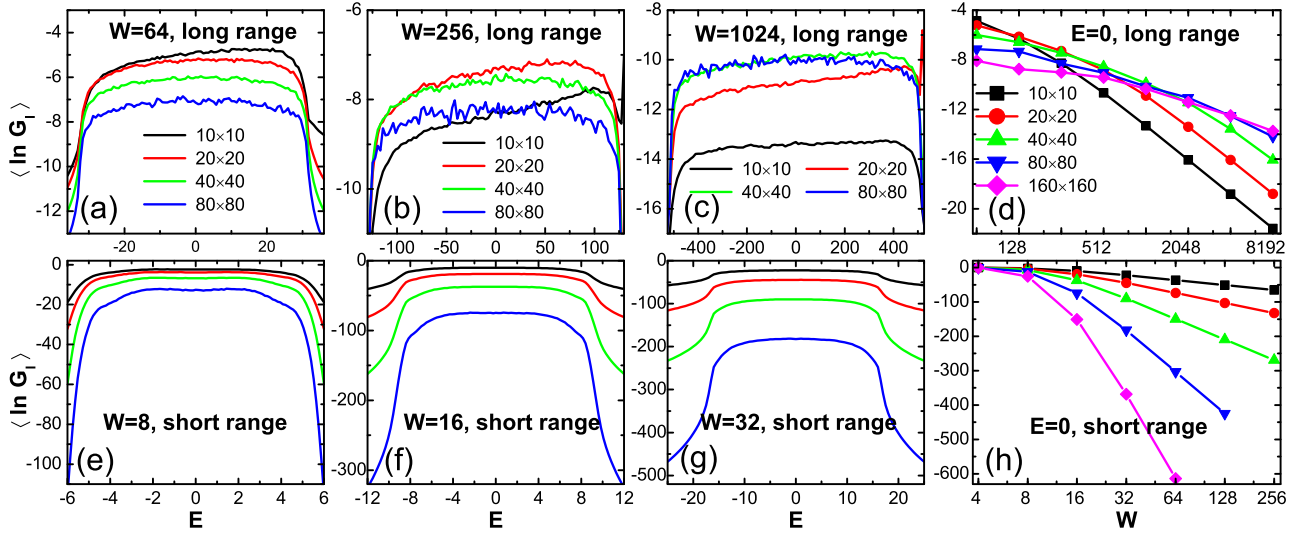}
\caption{(Color online) Disorder averaged $\langle \ln G_{\mathrm{I}}\rangle$ for the long range model (upper row) and the short range model (lower row). Left three columns: $\langle \ln G_{\mathrm{I}}\rangle$ versus Fermi energy $E$ for different disorder strengths $W$ . Rightmost column: $\langle \ln G_{\mathrm{I}}\rangle$ versus disorder strength $W$ (logarithmic scale) at Fermi energy $E=0$.  Curves in different colors correspond to different sample sizes: $10 \times 10 $ (black), $20 \times 20 $ (red), $40 \times 40 $ (green), $80 \times 80 $ (blue), and $160 \times 160$ (magenta). The number of disorder configurations for averaging is 1000 for $80 \times 80 $ and
$160 \times 160 $, and 5000 for other sizes. }
\label{FigScalingV2}
\end{figure}

Now we discuss this question: in the presence of disorder, will the long range hopping lead to a really metallic (delocalized) state in the thermodynamic limit? To answer this, one needs to perform size scaling on some transport quantities. A commonly used scaling quantity is the localization length normalized by the sample size, which can be extracted from the transfer matrix method\cite{TransferMatrix1,TransferMatrix2}. However, a transfer matrix can only be applicable to a model with a very finite hopping range. Here instead, we use the numerical scaling on the intrinsic conductance $G_{\mathrm{I}}$ which is defined as\cite{Braun1997}
\begin{equation}
\frac{1}{G_\mathrm{I}}=\frac{1}{G}-\frac{1}{M_c},\label{EqGI}
\end{equation}
with $M_c$ the number of active channels at Fermi energy in the lead and $G$ is the above used Landauer conductance.
The second term $\frac{1}{M_c}$ is used to deduct the effect of contact resistance at the sample-lead interfaces, so that $G_\mathrm{I}$ can manifest the intrinsic transport property of the bulk sample, which is found to be closely related to the conductance derived from the Kubo formula or Thouless formula \cite{Braun1997}. When $G$ is small (strong disorder) and/or $M_c$ is large (large size), $G_\mathrm{I} \thickapprox G$. This intrinsic conductance ($N_x \times N_x$) has been widely employed to investigate the occurrence, scaling and critical properties of the MIT in 2D and 3D systems\cite{Braun1997,Slevin2001,LocalizationGraphene,XRWang2019,HHWang2023}. For example, it can be used to evaluate the standard scaling function $\beta =\frac{d \langle\ln G_\mathrm{I}\rangle }{d \ln N_x}$ \cite{Abrahams79,AdersonTransitionReview1,AdersonTransitionReview2} of MIT, where $\langle \cdots \rangle$ still stands for averaging over the disorder ensemble. An increase (decrease) of $ \ln G_{\mathrm{I}} $ with increasing $\ln N_x$ indicates a metal with $\beta>0$ (insulator with $\beta<0$) phase\cite{Slevin2001,ConductanceExperiment}. In fact experimentally, scaling of the conductance is also the standard method to distinguish metal and insulator phases of materials\cite{MITFM,Gasparov2012,Givan2012,MITExp2016A,MITExp2016B,MITExp2016C}.

Again let us first take a look at the short range model that has been well understood, results of which are displayed in the lower rows of Figure \ref{FigScalingV2}. In panels (e), (f) and (g), $\langle\ln G_\mathrm{I}\rangle$ as a function of the energy $E$ is plotted for different disorder strengths, and panel (h) is $\langle\ln G_\mathrm{I}\rangle$ versus $W$ at a fixed energy $E=0$.  In each panel, curves in different colors correspond to results from different sample sizes. Here, the conductance is monotonically decreasing with the sample size and disorder strength for all states, which means that they are trivially
localized. Moreover, this decrease is more rapid for a larger sample size or a larger disorder strength.

On the other hand, in the upper row of Figure \ref{FigScalingV2}, the results for the long range model are quite different.
The first obvious feature is that for a definite size and disorder strength, the conductance of the long range model is several orders larger than that of the short range model.
For $W=64$ shown in Figure \ref{FigScalingV2} (a), $\ln G_\mathrm{I}$ is decreasing with increasing the sample size, in the whole energy region. Such decreasing can also be seen for weaker disorder strengths shown in the Supplementary Material.
These seem to indicate that all states of the long range model are localized in the thermodynamic limit. However, for a larger disorder $W=256$ as shown in Figure \ref{FigScalingV2} (b),
the conductance has a significant increase when the size is increased from $N_x=10$ (black) to $N_x=20$ (red) before a decrease again with larger sizes. For an even larger disorder $W=1024$ presented in Figure \ref{FigScalingV2} (c), the conductance increases up to size $N_x=40$ and then saturate there when the size is doubled as $N_x=80$.

In Figure \ref{FigScalingV2} (d), we fix the fermi energy at $E=0$ and display the development of the size scaling with increasing the disorder strength $W$. From the first glance, it seems to be a scaling pattern for the MIT, where the conductance is increasing ($\beta>0$) or decreasing ($\beta<0$) with $N_x$ on two ends of the $W$ axis, respectively\cite{Braun1997,Slevin2001,XRWang2019,HHWang2023}. However a careful scrutinize can reveal several distinct differences from the scaling pattern of the standard MIT. First, here curves associated with different sizes do \emph{not} cross at a single critical value $W_c$. Instead, curves for larger sizes cross at larger $W$. Second, on the larger $W$ side, although the conductance is clearly increasing with the scaling at small sizes, this increase eventually tends to saturate at large sizes.
For example, at $W=4096$, although the conductance is still increasing when the size is doubled from $40\times 40$ (green) to $80\times 80$ (blue), it saturates when the size is doubled again to $160\times 160$ (magenta). Therefore at $W=8192$, although the conductance of size $160\times 160$ is slightly larger than that of $80\times 80$, it is reasonable to imagine it will also saturate at, say, size $320 \times 320$ (which is beyond our calculation capability). Third, with increasing size,
on the side of increasing or saturating conductance (large $W$), the magnitude of the conductance is smaller than that on the side of decreasing conductance (small $W$). This is contrary to the case of the standard disorder induced MIT\cite{StarModel}, where the conductance of the metal phase (increasing conductance) is surely larger than that of the insulator phase (decreasing  conductance). In one word, at strong disorder, the transport property of this 2D long range model is neither a typical insulating phase ($\beta>0$) nor a typical metal phase ($\beta>0$), but seems to be marginal phase ($\beta\simeq 0$) which is a mixture of both. This is consistent with the physics shown in the upper row of Figure \ref{FigGammaEig}, where the disorder destroys the fully extended nature of the wavefunction ($\Gamma \ll 2$), but meanwhile the long range hopping prevents the wavefunction from a complete localization ($\Gamma \simeq 0$). Such a marginal state persists over several orders of $W$, typically with a constant fractal dimension $\Gamma$ [Figure \ref{FigGammaEig} (c)], or a slowly decreasing conductance $G_{\mathrm{I}}$ [Figure \ref{FigScalingV2} (d)].

Before closing this section, let us add some remarks on $W$ dependencies in Figure \ref{FigGammaEig} (c) and Figure \ref{FigScalingV2} (d). In the former figure, the typical fractal dimension of wavefunctions $\Gamma_{\mathrm{typ}}$ is almost constant when the disorder strength $W$ increases from 128 to 4096, which means the extent of the wavefunction's localization is \emph{independent} of the $W$. However in the latter figure, the conductance $G_{\mathrm{I}}$ is clearly \emph{decreasing} with $W$. There seems to be a discrepancy: is the transport independent of or decreasing with the disorder strength? To answer this, we need to remember that the conductance is a contribution from \emph{all} states around the Fermi energy\cite{Datta,Datta2}. As a result, the conductance is determined both by the ``diffusion coefficient'' (which can be characterized by the localization extent in real space or the level curvature in momentum space) of each wavefunction, and also by the number of wavefunctions (i.e., the density of states) round the Fermi energy\cite{Braun1997}.
Although Figure \ref{FigGammaEig} (c) states that increasing $W$ does not change the localization extent of each state,
it reduces the density of states by broadening the energy band. This leads to a reduction of the conductance with increasing $W$ displayed in Figure \ref{FigScalingV2} (d).

\section*{The Isolated State}

One should remember that there is a single state (\ref{EqIsolatedState}) isolated far away ($\sim N^2_x \gamma $) from the band investigated above. In 1D with periodic boundary condition, it was found that the wavefunction of this isolated state is fully extended with a uniform magnitude and phase among all lattice sites\cite{Shielding2016}. In 2D and in the fixed boundary condition here, we have checked (but do not show here) that this uniformity is only slightly changed, and the energy value of this state is also slightly different from that given in Equation (\ref{EqIsolatedState}).

\begin{figure}[htbp]
\centering
\includegraphics*[width=0.7\textwidth]{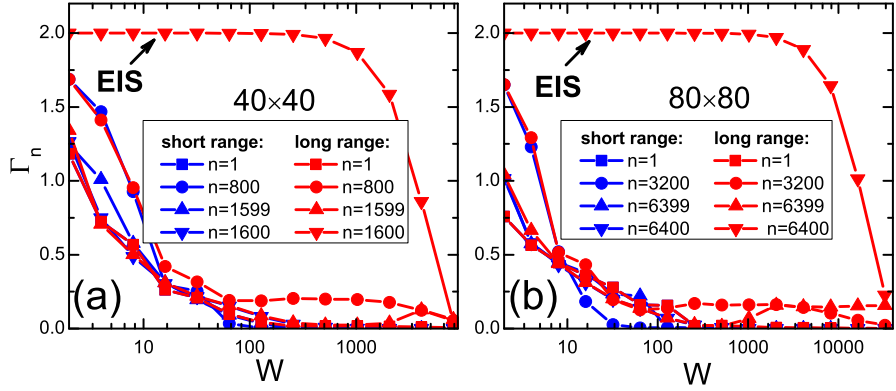}
\caption{(Color online) The fractal dimension $\Gamma_n$ of the $n$-th eigenstate as a function of disorder strength $W$, for a certain sample with size $40 \times 40$ (a), and size $80 \times 80$ (b). Blue (red) color is for the short (long) range model. Notice horizontal scales in two panels are different. }
\label{FigGamma}
\end{figure}

As a first step, we also study the fractal dimension $\Gamma_n$ of eigenstates defined in equation \ref{EqFractalDimension}. In Figure \ref{FigGamma}, red symbol-curves are $\Gamma_n$ of four representative states as functions of disorder strength $W$: the band bottom (square), the band center (circle), the band top (up triangle), and the isolated state (down triangle). Panels (a) and (b) are for a $40\times 40$ and a $80\times 80$ sample respectively. For comparison, results from a short range model are also shown as blue symbol curves. We can see that for eigenstates in the band, $\Gamma_n$ decays away from 2 rapidly with increasing disorder.

On the contrary, the fractal dimension for the isolated state is robustly quantized as 2 until $W$ approaches $\sim N^2_x \gamma $, the gap from the band. We remind that it is a \emph{single} state instead of a flat band consisting of a continuum of \emph{states}. Thus for an electron in this state, there is \emph{no} other state for it to be scattered onto, and therefore any back-scattering or skew-scattering
is forbidden. The robustness of this state's transport is protected by the large gap from the band, unless the impurity strength is strong enough $\sim N^2_x \gamma$ to overcome this large gap. This is the physical origin of the robust transport of this state, which we call it the extended isolated state (EIS) hereafter.

\begin{figure}[htbp]
\centering
\includegraphics*[width=0.65\textwidth,bb=0 245 486 597]{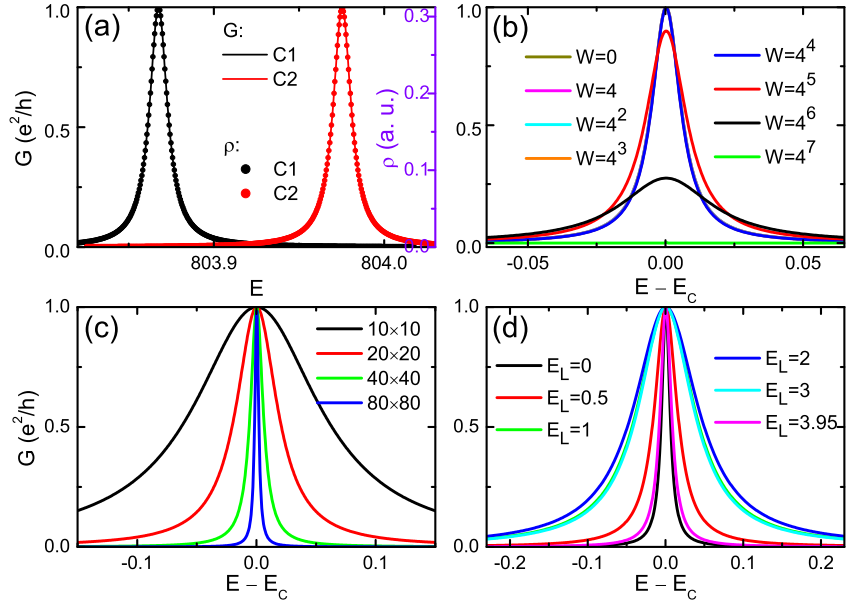}
\caption{(Color online) Two-terminal conductance $G$ as a function of the Fermi energy in the sample, for the extended isolated state. (a) For two disorder configurations, C1 (black curve) and C2 (red curve). The corresponding DOS $\rho$ (referenced to the right axis) are also plotted as dots. (b) For different disorder strengths $W$. (c) For different sample sizes. (d) With different energies $E_L$ in leads. If not otherwise stated in the panel, parameters are as follows: $W=16$, $E_L=0$ and size $40\times 40$. In panels (b), (c) and (d), all curves have been horizontally shifted to the same peak center. }
\label{FigIsolatedAll}
\end{figure}

The above results are from a sample decoupling from the environment. Now we still simulate the quantum transports of the EIS by attaching leads to both ends of the sample. Here, since the condition Equation (\ref{EqCommute}) for perfect shielding is partially destroyed by the fixed boundary condition, we have checked that this EIS is not an eigenstate of the short range model (played as leads). In other words, no state in the lead can actually match the EIS in the central sample. Therefore in this section, when calculating transports by using Eqs. (\ref{EqConductance}) and (\ref{EqLocalCurrent}), as illustrated in Figure \ref{FigIllustration} (c), we simply treat the leads as electron reservoirs with a fixed Fermi energy $E_L$ at the band center and vary the Fermi energy of the sample $E$ around the EIS, e.g., $G(E)=\frac{e^{2}}{h}\mathrm{Tr}\left[  \Gamma_{L}(E_L)G^{r}(E)\Gamma
_{R}(E_L)G^{a}(E)\right]$.

In Figure \ref{FigIsolatedAll} (a), conductances of two disorder configurations [C1 (black curve) and C2 (red curve)] as functions of $E$ are presented as solid curves, with size $40 \times 40$ and disorder strength $W=16$. The corresponding density of states (DOS)
\begin{equation}
\rho(E)= -\frac{1}{\pi N_x^2} \mathrm{Im} G^r(E),\label{EqDOS}
\end{equation}
are also plotted as dots. Due to the coupling with leads, both the conductance and DOS profiles are broadened as smooth peaks. The first obvious observation is the perfect coincidence between profiles of $G(E)$ and $\rho(E)$, after appropriate scaling in the vertical direction. This can be understood as follows. The zero-temperature conductance of a 2D crystal can also be expressed in the Thouless form \cite{Thouless,Braun1997},
\begin{equation}
G=\pi\rho(E)\frac{\partial^2 E}{\partial k_x^2}\Big|_{k_x=0}, \label{EqThouless}
\end{equation}
where $k_x$ is the wavevector along the transport direction. Notice the last factor $\frac{\partial^2 E}{\partial k_x^2}$ is a curvature of the band $E(\bm{k})$. However here, there is only a single state for which no curvature can be defined, so this factor plays no role and thus has no energy dependence. As a result for a concrete sample, the energy dependence of the conductance is simply proportional to that of the DOS, both with the same spectrum width determined by the imaginary part of the $N$-th eigenvalue of the dressed (non-Hermitian) Hamiltonian $H+\Sigma_L+\Sigma_R$.

The second observation from Figure \ref{FigIsolatedAll} (a) is more important. The peak value of the conductance is always quantized as unity, which means the EIS carries a perfectly conducting channel. This reflects the absence of backscattering for the EIS, as predicted from Figure \ref{FigGamma} above. Conductance peaks with other $W$ are shown in Figure \ref{FigIsolatedAll} (b). The robust transport of EIS persists at least to $W=4^4=256$ for the sample size $40 \times 40$, consistent with Figure \ref{FigGamma} (a). To reveal the microscopic origin of this robust transport, we again turn to the real space distribution of bond currents at the conductance peak, for a smaller sample as shown in Figure \ref{FigCurrent} (f). The picture is simple and clear: the dominating currents are flowing through very long bonds connecting the left and right boundaries \emph{directly}. This configuration helps the electron to circumvent any impurities in the sample bulk, leading to a remarkably robust transport independent of disorder.

The size dependence of the conductance peak is presented in Figure \ref{FigIsolatedAll} (c), where larger samples result in sharper conductance peaks. This is not surprising because a larger size leads to a finer resolution of energy and thus a smaller broadening. Figure \ref{FigIsolatedAll} (d) are conductance peaks with different lead energies $E_L$ (respect to its band center). It is interesting to notice that sharpest peaks correspond to injecting electrons from the band edge ($E=3.95$) or the band center ($E\sim 0$) of the leads.

\section*{Summary}
We numerically investigate the quantum transports of a 2D system with long range hopping. In the band, the transport is almost identical to that in the corresponding short range system for a large sample size and weak disorder, as a manifestation of the cooperative shielding. This shielding is broken at strong disorder, and the average and fluctuation of the conductance are larger than those in the short range mode. These can be understood as a better connectivity and the destruction of path pinning from long range coupling. Over several orders of strong disorder, the band states exhibit a marginal feature between metallic and insulating states, which is neither fully extended nor completely localized. As for the isolated state, its transport of a unit conducting channel is highly robust against disorder, which can be depicted as an extended state protected by the large gap from scattering.

\section*{Data availability}
On reasonable request, the corresponding author will provide all relevant data in this paper.

\section*{Code availability}
On reasonable request, the corresponding author will provide all numerical codes in this paper.


\section*{Acknowledgements}

This work was supported by Guangdong Basic and Applied Basic Research Foundation under No. 2023A1515010698, National Natural Science
Foundation of China under Grant Nos. 12104108 and 11874127, the Joint Fund with Guangzhou Municipality under Nos. 202201020198 and 202201020137, and the Starting Research Fund from Guangzhou University under Grant Nos. RQ2020082, RQ 2020083 and 62104360.

\section*{Author contributions statement}

S.-S. Wang, Y.-M. Dai and Y.-Y.Z. carried out the theoretical calculations and wrote the manuscript with the assistance of K. Li, H.-H. Wang and Y.-C. Zhang. Y.-Y. Zhang and Y.-C. Zhang guided the overall project. All authors reviewed the manuscript.

\section*{Additional information}

\textbf{Competing financial interests:} The authors declare no competing financial interests.

\newpage

\renewcommand{\theequation}{S\arabic{equation}}
\renewcommand{\thefigure}{S\arabic{figure}}
\setcounter{equation}{0}
\setcounter{figure}{0}

\section*{Supplementary Material on Quantum Transports in Two-Dimensions with Long Range Hopping}

  This supplementary material shows the scaling of conductance for the 1D long range model.

\begin{figure}[htbp]
\centering
\includegraphics*[width=0.6\textwidth,bb=0 254 481 600]{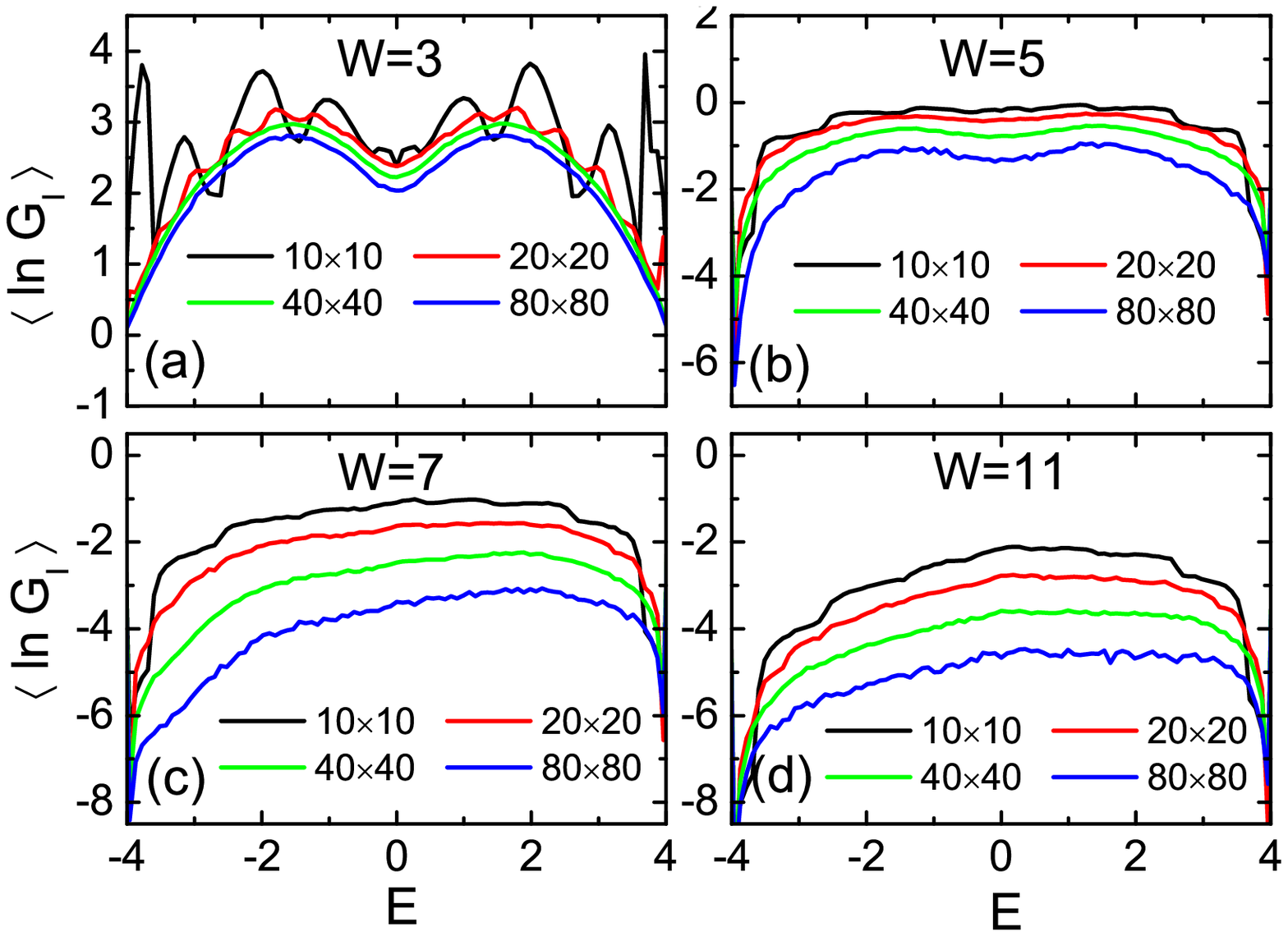}
\caption{(Color online) Disorder averaged $\langle \ln G_{\mathrm{I}}\rangle$ as a function of Fermi energy $E$ for different sample sizes: $10 \times 10 $ (black), $20 \times 20 $ (red), $40 \times 40 $ (green), $80 \times 80 $ (blue), at different disorder strengths: (a) $W=3$, (b) $W=5$, (c) $W=7$, (d) $W=11$. The average is over 1000 samples for the largest size and 5000 samples for other sizes. }
\label{FigScaling}
\end{figure}

In this Supplementary Material, we display the scaling results for weak disorder. In the extremely weak disorder regime ($W=3$)  shown in Figure \ref{FigScaling} (a), there are significant finite-size fluctuations for the smallest size ($10\times 10$, black curve). After they are smoothed out in large sizes, it is clear that $\ln G_\mathrm{I}$ is decreasing with increasing sample size, in the whole energy region. This decreasing is more significant at larger disorder strengths, as shown in the rest panels of Figure \ref{FigScaling}, indicating a strong trend of localization. There seem to be some curve crossings around two ends of the energy interval. We have checked that this only occurs for the smallest size $10 \times 10$, and that for larger sizes, $\ln G_\mathrm{I}$ is still monotonically decreasing. Therefore, within the best numerical capability we can achieve, all band states with $W\geqslant 3$ is shown to be localized.

\end{document}